\documentclass[preprint,showpacs,preprintnumbers,amsmath,amssymb,superscriptaddress]{revtex4}

\usepackage{graphicx}
\usepackage{dcolumn}
\usepackage{amsmath}    
\usepackage{verbatim}   
\usepackage{color}      
\usepackage{subfigure}  
\usepackage{hyperref}   
\usepackage{bm}

\begin{document}


\title{Tunable Indistinguishable Photons From Remote Quantum Dots
}

\author{R. B. Patel}
\affiliation{Toshiba Research Europe Limited, Cambridge Research Laboratory,\\
208 Science Park, Milton Road, Cambridge, CB4 OGZ, U. K.}
\affiliation{Cavendish Laboratory, Cambridge University,\\
JJ Thomson Avenue, Cambridge, CB3 0HE, U. K.}
\affiliation{These authors contributed equally}

\author{A. J. Bennett}
\email{anthony.bennett@crl.toshiba.co.uk}
\affiliation{Toshiba Research Europe Limited, Cambridge Research Laboratory,\\
208 Science Park, Milton Road, Cambridge, CB4 OGZ, U. K.}
\affiliation{These authors contributed
equally}

\author{I. Farrer}
\affiliation{Cavendish Laboratory, Cambridge University,\\
JJ Thomson Avenue, Cambridge, CB3 0HE, U. K.}

\author{C. A. Nicoll}
\affiliation{Cavendish Laboratory, Cambridge University,\\
JJ Thomson Avenue, Cambridge, CB3 0HE, U. K.}

\author{D. A. Ritchie}
\affiliation{Cavendish Laboratory, Cambridge University,\\
JJ Thomson Avenue, Cambridge, CB3 0HE, U. K.}

\author{A. J. Shields}
\affiliation{Toshiba Research Europe Limited, Cambridge Research Laboratory,\\
208 Science Park, Milton Road, Cambridge, CB4 OGZ, U. K.}

\date{\today}%

\begin{abstract}
Single semiconductor quantum dots have been widely studied within devices that can apply an
electric field. In the most common system, the low energy offset between the InGaAs quantum dot and
the surrounding GaAs material limits the magnitude of field that can be applied to tens of
kVcm$^{-1}$, before carriers tunnel out of the dot. The Stark shift experienced by the emission
line is typically $\sim1\textrm{ meV}$.  We report that by embedding the quantum dots in a quantum
well heterostructure the vertical field that can be applied is increased by over an order of
magnitude whilst preserving the narrow linewidths, high internal quantum efficiencies and familiar
emission spectra. Individual dots can then be continuously tuned to the same energy allowing for
two-photon interference between remote, independent, quantum dots.
\end{abstract}


\maketitle 

In recent years self-assembled semiconductor quantum dots have been established as one of the most
versatile systems for studying quantum effects in the solid-state as well as quantum optics. A
plethora of quantum dot based field-effect devices have been developed which allow controlled
charging of a quantum dot\cite{warburton00}, Rabi oscillations\cite{zrenner02}, coherent spin
control\cite{atature06} and electrically injected non-classical photon emission\cite{yuan02}. Often
referred to as ``artificial atoms", quantum dots possess discrete energy levels which make them a
viable candidate for qubits. However, unlike single atoms, no two quantum dots are alike and
experiments are often limited due to the distributions in energy, fine-structure and linewidths in
which they form. This is a complication for quantum information applications that require qubits to
be initialized in the same state. Furthermore, distributed quantum computing requires interactions
between remote qubit systems mediated by indistinguishable photons. Quantum effects between remote
sources have been observed with ions\cite{maunz07,olmschenk09}, atoms\cite{beugnon06,chaneliere07}
and parametric down-conversion\cite{kaltenbaek06,halder07} in which photons are naturally generated
with the same energy.  Recently, quantum interference has been demonstrated between donor
impurities on the same chip\cite{sanaka09}. Though in this case the visibility of interference is
hampered by a lack of control over the emission energy and unfavorable multi-photon emission. It
would be desirable to be able to precisely engineer the characteristics of each emitter
independently.

Consider the probability of finding two quantum dots with the same transition energy.  Typically,
the distribution of energies in a sample can be as low as 10 meV, however the narrowest linewidths
tend to range from $1-5 \textrm{ }\mu{\textrm{eV}}$\cite{patel08,michler09}. Therefore there is a
minute chance of finding two dots with perfectly overlapping spectra.  Perhaps the easiest method
of tuning the transition energy is by applying a vertical electric field, however, tunneling of
carriers out of the dot results in a Stark shift of only $\sim1\textrm{ meV}$ before the emission
is quenched\cite{finley02,findeis01}.  We circumvent this issue by embedding our quantum dots in
the center of a GaAs/AlGaAs quantum well.  Not only does this retain the favorable properties of
the InGaAs/GaAs quantum dot system, but it also allows carrier confinement even in the presence of
large vertical electric fields leading to energy shifts of up to 25 meV. We demonstrate these
states are quantum mechanically identical by carrying out two-photon interference between photons
emitted by remote quantum dots.

To begin with, we shall briefly discuss our samples.  The first, which we denote Dot A, constitutes
our reference sample and is an electrically injected InAs quantum dot embedded in a microcavity
p-i-n diode.  Previous studies of this sample have shown a low multi-photon emission probability,
narrow linewidth and bright electroluminescence (EL).  The emitting state is a negatively charged
exciton $\left(X^{-}\right)$ emitting at $1.314 \textrm{ eV}$.  This state has exhibited a high
degree of indistinguishability between consecutively emitted photons and also with a coherent
laser\cite{bennett09}.  The second set of samples consist of a single layer of InAs quantum dots
embedded in the center of a 10 nm GaAs quantum well clad with
$\textrm{Al}_{0.75}\textrm{Ga}_{0.25}\textrm{As}$.  These layers encompass a half-wavelength thick
cavity between two Bragg reflectors.  The microcavity serves to enhance the photon collection
efficiency with negligible effect on the emission rate.  The indium deposition and layer
thicknesses of the microcavity were tailored for emission at around $1.314 \textrm{ eV}$.  Doping
was introduced allowing a vertical electric field to be applied.  A schematic of the band structure
is shown in Fig. 1a.  Single dots were isolated using an array of apertures patterned in the top
contact of the device.  Common to nearly all single quantum dots is the change in the carrier
occupancy as a function of effective field $F$ (Fig. 1b).  Exciton $(X)$, biexciton $(XX)$ and
charged exciton $(X^{-}\textrm{, }X^{+})$ states can be identified.  Each state shifts in a
parabolic fashion with fields up to $-500 \textrm{ kVcm}^{-1}$. This is an order of magnitude
increase over previously reported values where $60 \textrm{ kVcm}^{-1}$ fields would result in
quenching of the emission\cite{finley02,findeis01}. Stark shifts of up to 25 meV have been observed
for single $X$ states, with measured polarizabilities and dipole moments consistent with data
reported with much smaller fields\cite{finley04}.  Although some selectivity is applied to identify
the brightest and narrowest lines we find most dots can be tuned to the target energy of $1.314
\textrm{ eV}$. Below we utilize the $X^-$ states of the dots.

 In Fig. 1c we demonstrate tuning of
two quantum dot $X^-$ states to the same energy.  In this case, the tunable dot (Dot 1) emits at
the target energy of $1.314 \textrm{ eV}$ with a net field of $-50 \textrm{ kVcm}^{-1}$.
High-resolution spectroscopy (Fig. 1d) with a scanning Fabry-P\'{e}rot interferometer ($0.8
\textrm{ }\mu\textrm{eV}$ resolution) allows us to tune the two states to degeneracy.   Dot A,
reveals a Lorentzian lineshape with a FWHM of $5.2\textrm{ }\mu{eV}$.  Dot 1 however is best fitted
with a Voigt profile with a homogenous linewidth of $2.2\textrm{ }\mu{\textrm{eV}}$ and a Gaussian
width of $6.8\textrm{ }\mu{\textrm{eV}}$.  We note that most lines in other dots exhibit a
predominantly Lorentzian lineshape.

To further demonstrate the control we have over the energy, we now discuss two-photon interference
measurements taken with these two dots. Two-photon interference refers to the phenomena whereby
single photons entering a 50:50 beamsplitter from opposite inputs ``coalesce" and exit together,
provided they are indistinguishable\cite{hom87}.  This is also used as a test of purity of a
quantum light source\cite{santori02,patel08,bennett08}. Both samples were held in two separate
cryostats located $1.1\textrm{ }\textrm{ m}$ apart at $4.5 \textrm{ K}$.  For Dot A the device was
held at a fixed bias corresponding to a current of $130 \textrm{ }\mu{A}$, whilst Dot 1 was excited
with a continuous-wave laser emitting at $1.459 \textrm{ eV}$.  These conditions were chosen to
give equal intensity, bright emission from each sample whilst also maintaining narrow linewidths.
Emission from each dot is directed into a polarizing beamsplitter (Fig. 2) where the orthogonally
polarized, co-parallel, beams are spectrally filtered using a monochromator.  They are then coupled
into a custom-built Mach-Zehnder interferometer made out of polarization-maintaining single-mode
fiber.  A polarizing coupler separates the emission from each source and sends them along
individual paths onto a 50:50 coupler. We periodically rotate the paths from being mutually
parallel in polarization (indistinguishable) or orthogonal in polarization (distinguishable).  We
thereby build correlations for both cases during the same experimental run, canceling any effects
due to drift. Photons are detected using two silicon avalanche photodiode detectors, $D_{1}$ and
$D_{2}$.  The temporal response of the system $R_{f}$ is a Gaussian with a FWHM of $428\textrm{
ps}$.

 First we determine the multi-photon emission probability for each source.  These were
measured in a Hanbury-Brown and Twiss experiment by blocking the emission from one of the sources
whilst measuring the second-order correlation for the other.  The data in Fig. 3a and b were fitted
with the function $g^{(2)}(\tau) = R_{f}\otimes(1-(1-B)\textrm{exp}(-|\tau|/\tau_{r})))$ where
$\tau$ is the time between detections, $B$ is the background contribution and $\tau_{r}$ is the
radiative lifetime of the source.  The blue curves in each figure are fits assuming an infinitely
fast response, $B = 0.05$ and $\tau_{r} = 600\textrm{ ps}$ for Dot A, and $B = 0$ and $\tau_{r} =
800\textrm{ ps}$ for Dot 1.  Convolving with the known response of our detectors (red curve)
results in an excellent fit to the data.

In Fig. 3c and d we present interference data where the energy detuning of Dot 1 from Dot A,
$\Delta{E} = 0\textrm{ }\mu{\textrm{eV}}$, and each source delivers photons with parallel
polarization and orthogonal polarization respectively.  The signature of two-photon interference is
a difference in the depth of the dip at $\tau=0\textrm{ ns}$. For orthogonally polarized photons,
the photons are distinguishable and no interference occurs. Therefore, there is equal chance of the
two photons exiting the beamsplitter in the same direction or in opposite directions result in a
dip to 50\%. Indistinguishable photons with parallel polarization give rise to interference leading
to a suppression in coincident counts and a dip to 0\%. For each measurement the dip does not reach
the ideal values of 0\% and 50\% respectively, due to the response of our detectors.  We have
developed a model to account for this and also the difference coherence times, beamsplitter
coefficients, imperfect wavepacket overlap $\gamma=\left\langle\psi_{A}|\psi_{1}\right\rangle$, and
finite multi-photon emission $g^{(2)}(0)$ for each source. We also account for the Gaussian
component of Dot 1 numerically by including a spectral jitter in the emission energy whose
magnitude is determined from a Voigt fit of the spectra. As a measure of mutual
indistinguishability between the two sources, we define the visibility of two-photon interference
as $V = (g^{(2)}_{\perp}(\tau)-g^{(2)}_{\parallel}(\tau))/g^{(2)}_{\perp}(\tau)$.  By
post-selecting events where photons arrive simultaneously at the final coupler, we measure a
visibility of $33 \pm 1\%$ from the raw data (Fig. 3e) which agrees with our model. As shown by the
blue curve, without the limitation of timing resolution it should be possible to achieve a
visibility of up to $98\%$.

Scaling a system to multiple qubits is an important requirement for realizing many useful
applications in quantum information processing. Linear optical schemes using quantum dots would
greatly benefit from the ability to interfere photons from arrays of quantum dots in separate
locations. Our technique can be used to tune a number of quantum dots to the same energy. We
demonstrate this with our devices by repeating the interference measurements for two other dots.
High-resolution PL measurements of Dot 2 and 3 indicate a Lorentzian lineshape and negligible
inhomogeneous broadening. Table 1 provides a summary of these results where in each case we infer
an overlap close to unity. This shows that we can observe interference with a visibility that can
be determined from independent Hanbury-Brown and Twiss and coherence time measurements. This
ability to produce many indistinguishable quantum states in remote solid-state systems is required
for quantum communication networks and distributed quantum computing.

 Control over the emission
energy allows us to perform an experiment analogous to the famous ``Hong-Ou-Mandel dip" experiment
in which a rate of coincidences is recorded as some distinguishing character is introduced between
photons.  Conventionally, this is achieved by delaying a photon in time relative to
another\cite{hom87,santori02,bennett08}. In our experiment we introduce a controllable energy
difference between Dot A and Dot 1 and measure the visibility of interference which is shown in
Fig. 3f. Notice that a high degree of interference is only seen when $\Delta{E}$ is close to zero.
For two sources that are homogenously broadened, it can be shown that the width of the peak in Fig.
3f should be equal to the sum of the linewidths of the two sources (in the limit where the
coherence time is much less than the radiative lifetime and detection resolution). Our model
predicts a similar width of $15\textrm{ }\mu{\textrm{eV}}$. In contrast, the experiment reveals a
peak of width $4.9\textrm{ }\mu{\textrm{eV}}$. This suggests that for finite detunings the sources
are more distinguishable than is suggested by their spectra. This would be the case if information
about the emitted photons were retained by the sources after emission, for instance if a given
photon had a narrower linewidth than that observed in the time-averaged spectra. Such an
observation is outside the scope of the accepted models of two-photon interference\cite{legero04}
but we hypothesize that this information is retained by the solid-state environment around the
quantum dot after radiative recombination from the $X^-$ state, for example in the fluctuations
that lead to inhomogeneous broadening of the dot. An interesting way to test this would be to probe
the environment directly using a single shot absorption technique immediately after the photon is
emitted. In future higher visibilities of interference could be achieved using
pulsed\cite{bennett08} or resonant excitation\cite{michler09}.

The ability to apply large electric fields to single quantum dots, whilst maintaining high quantum
efficiencies, provides a major step towards refining quantum dot based single-photon sources.  The
high degree of control over the transition energy we have shown solves the long standing problem of
identifying quantum dots with the same energy. Thus allowing, for the first time, two-photon
interference between  remote quantum dots. Experimental measurements of the interference visibility
are in agreement with our predictions for the three pairs of quantum dots studied showing that a
large degree of overlap is achievable between photons emanating from different dots. This work
opens up the possibility of transferring quantum information between multiple, remote, sources in
the solid-state.

This work was partly supported by the EU through the IST FP6 Integrated Project Qubit Applications
(QAP: contract number 015848). EPSRC provided support for RBP and QIPIRC for CAN.

    \begin{center}
    \begin{figure}[h]
    \includegraphics[width=100mm]{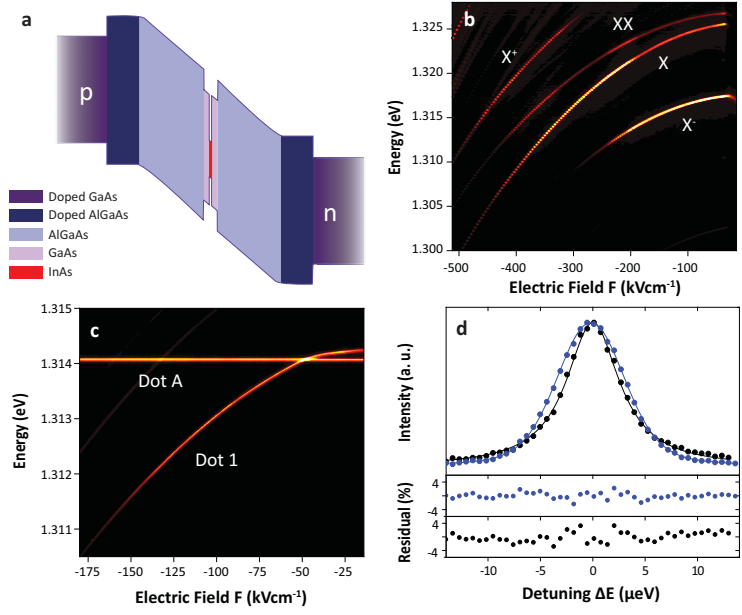}

    \caption{\label{Fig1} Design and spectral charactersitics of our tunable source.
    (\textbf{a}) Schematic illustrating the band structure near the active region of the tunable
    device. A layer of InAs quantum dots is grown at the center of a 10 nm GaAs quantum well clad with
    a $\textrm{Al}_{0.75}\textrm{Ga}_{0.25}\textrm{As}$ short period superlattice. (\textbf{b}) Typical
    spectra of a single quantum dot as the field is varied.  The states are identified as exciton
    $(X)$, biexciton $(XX)$ and charged exciton $(X^{-}\textrm{, }X^{+})$. (\textbf{c}) Tuning Dot A
    and Dot 1 to the same energy. (\textbf{d}) High-resolution spectra of Dot A (black circles) and Dot
    1 (blue circles) at zero detuning. A Lorentzian spectrum (black curve) is observed for Dot A with a
    linewidth of $5.2\textrm{ }\mu{eV}$. Dot 1 shows a Gaussian component and is fitted with a Voigt
    profile (blue curve). Also shown are the residuals of the least-squares fit.}\end{figure}
    \clearpage
    \end{center}

    \begin{center}
    \begin{figure}[h]
    \includegraphics[width=120mm]{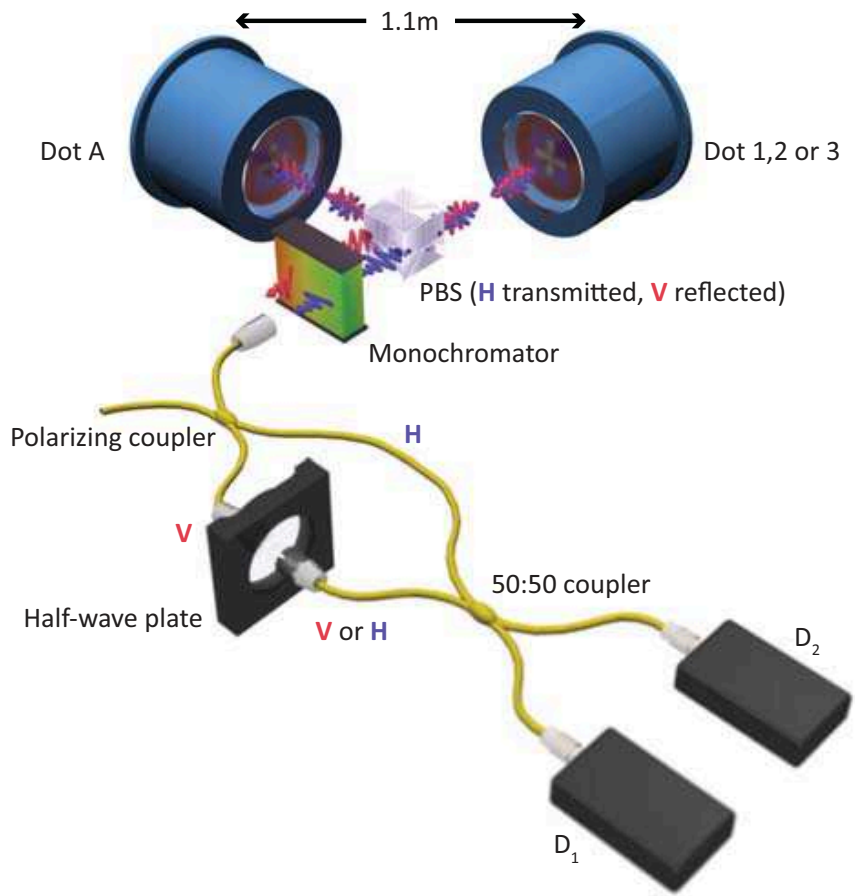}
    \caption{\label{Fig2} Experimental arrangement for two-photon interference.  The two sources are
    located 1.1 m apart. V polarized photons from Dot A and H polarized photons from Dot 1,2 or 3 are
    collected and filtered using a monochromator. The emission is coupled into a Mach-Zehnder
    interferometer made from polarization-maintaining single-mode fiber. Emission from the individual
    sources is separated using a polarizing coupler and then brought together in a 50:50 coupler.
    Correlations are recorded whilst periodically switching a half-wave plate making the photons
    mutually indistinguishable or distinguishable.}\end{figure} \clearpage
    \end{center}

    \begin{center}
    \begin{figure}[h]
    \includegraphics[width=150mm]{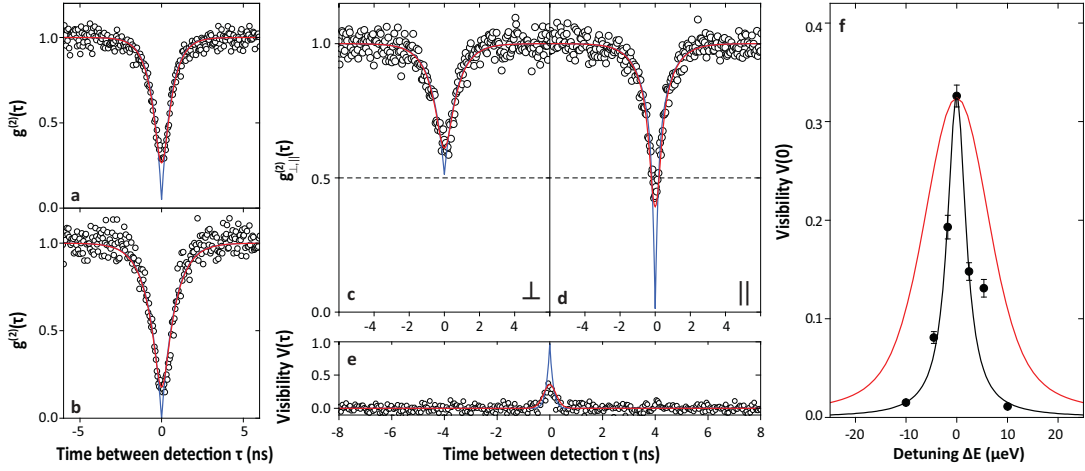}
    \caption{\label{Fig2}  Hanbury-Brown and Twiss and two-photon interference results. In
    (\textbf{a})-(\textbf{f}) blue curves correspond to fits assuming unlimited timing resolution and
    red curves to fits assuming a Gaussian response of $428 \textrm{ ps}$. (\textbf{a}) Autocorrelation
    for Dot A, with $g^{(2)}(0)= 5\%$. (\textbf{b}) Autocorrelation for Dot 1, with $g^{(2)}(0)= 0\%$.
    (\textbf{c-d}) Two-photon interference measurements with $\Delta{E} = 0\textrm{ }\mu{\textrm{eV}}$.
    (\textbf{c}) With mutually orthogonal polarized photons the dip is above the classical limit
    (dashed line) and no interference has occurred. (\textbf{d}) Parallel polarized photons result in a
    dip below the 50 \% owing to two-photon interference. (\textbf{e}) Raw interference visibility
    limited mainly by the timing resolution. (\textbf{f}) Measurements of visibility as a function of
    $\Delta{E}$ fitted with our model (red line) and with a Lorentzian with FWHM = $4.9\textrm{
    }\mu\textrm{eV}$ (black line)}
    \end{figure} \clearpage
    \end{center}

    \begin{center}
    \begin{figure}[h]
    \includegraphics[width=150mm]{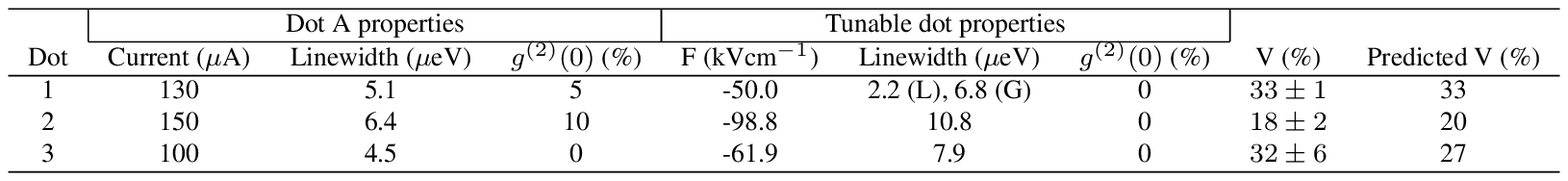}
    \caption{\label{Table1}  Summary of interference results for Dot A and 1-3. Labels L denotes lorentzian lineshapes, and G denotes Gaussian lineshapes.}
    \end{figure} \clearpage
    \end{center}


\end{document}